\documentclass[fleqn]{mnras}

\usepackage{newtxtext,newtxmath}

\usepackage[T1]{fontenc}
\usepackage{ae,aecompl}
\usepackage{epsf} 

\usepackage{graphicx}	
\usepackage{amsmath}	
\usepackage{amssymb}	
\usepackage{rotating}
\usepackage{caption}

\title[Anomalous behaviour of the UV--optical continuum bands in NGC~5548]{
Anomalous behaviour of the UV--optical continuum bands in NGC~5548}

\author[M.R. Goad et al.]{
M.~R.~Goad,$^{1}$\thanks{E-mail: mg159@le.ac.uk} 
C.~Knigge,$^{2}$
K.~T.~Korista,$^{3}$
E.~Cackett,$^{4}$
K.~Horne,$^{5}$ 
D.~A.~Starkey,$^{5}$\newauthor
B.~M.~Peterson,$^{6,7,8}$ 
G.~De~Rosa,$^{7}$,
G.~A.~Kriss,$^{7}$ 
R.~Edelson,$^{9}$
M.~Fausnaugh$^{6,10}$
\\
$^{1}$University of Leicester, Department of Physics and Astronomy, University Road, Leicester LE1 7RH, UK\\
$^{2}$School of Physics \& Astronomy, University of Southampton, Highfield, Southampton SO17 1BJ, UK\\
$^{3}$Western Michigan University, Department of Physics, 1120 Everett Tower, Kalamazoo, MI 49008-5252, USA\\
$^{4}$Wayne State University, Department of Physics and Astronomy, 666 W. Hancock St, Detroit, MI 48201, USA\\
$^{5}$SUPA Physics and Astronomy, University of St. Andrews, Fife, KY16 9SS Scotland, UK\\
$^{6}$Department of Astronomy, The Ohio State University, 140 W 18th Ave, Columbus, OH 43210, USA\\
$^{7}$Space Telescope Science Institute, 3700 San Martin Drive, Baltimore, MD 21218, USA\\ 
$^{8}$Center for Cosmology and AstroParticle Physics, The Ohio State University, 191 West Woodruff Ave, Columbus, OH 43210, USA\\
$^{9}$University of Maryland, Department of Astronomy, College Park, MD 20742-2421, USA\\
$^{10}$Department of Physics, Massachussetts Institute of Technology, 77 Massachussetts Avenue, Cambridge, MA 02139-4307, USA
}

\date{Accepted 2019 April 25. Received 2019 April 9; in original form 2018 December 8}

\pubyear{2019}

\begin{document}
\label{firstpage}
\pagerange{\pageref{firstpage}--\pageref{lastpage}}

\maketitle

\begin{abstract}
During the 2014 \textit{HST/Swift} and ground-based multi-wavelength
monitoring campaign of NGC~5548 (AGN STORM), the UV--optical broad
emission lines exhibited anomalous, decorrelated behaviour relative to
the far-UV continuum flux variability.  Here, we use key diagnostic
emission lines (Ly$\alpha$ and He~{\sc ii}) for this campaign to infer
a proxy for the all important, variable driving EUV continuum incident
upon BLR clouds. The inferred driving continuum provides a crucial
step towards the recovery of the broad emission line response
functions in this AGN. In particular, the ionising continuum seen by
the BLR was weaker and softer during the anomalous period than during
the first third of the campaign, and apparently less variable than
exhibited by the far-UV continuum. We also report the first evidence
for anomalous behaviour in the longer wavelength (relative to
$\lambda$1157\AA\/) \textit{continuum bands}. This is corroborative
evidence that a significant contribution to the variable UV--optical
continuum emission arises from a diffuse continuum emanating from the
same gas that emits the broad emission lines.

\end{abstract}


\section{Introduction}\label{section:intro}

Correlated continuum and broad emission line variability studies
(reverberation mapping, hereafter RM) have proven a powerful probe of
the central regions of Active Galactic Nuclei (AGN). Following the
seminal paper by Blandford \& McKee (1982), it was soon realised that
RM could not only reveal the spatial distribution and kinematics of
the line emitting gas, but also, with few assumptions, provide an
estimate for the mass of the central super-massive black hole (e.g.,
Peterson et al.\ 2002, 2014; Bentz et~al.\ 2009, 2010a,b, 2014; Grier
et~al.\ 2012; Kaspi et~al. 2002, 2007; Pancoast et~al.\ 2012,
2014a,b). This technique has yielded black hole mass estimates for
$\approx$60, mostly nearby and low-luminosity, AGN (see Bentz \& Katz
(2015), for details). RM black hole mass determinations are
competitive with mass estimates derived using more conventional
techniques (e.g., stellar velocity dispersion), while high source
luminosities afford access to mass determinations at high redshift. RM
is thus important for studying the accretion flow and energy
generation mechanism in AGN, and for investigating black hole growth
on cosmological timescales (Silk 2013; Silk \& Rees 1998).

While RM was largely developed for broad emission-line variability
studies, it has also been successfully applied to the continuum bands
alone. Specifically, correlated optical--IR continuum variations (dust
reverberation) can provide a robust estimate of the distance to the
hot-dust, an upper limit to the BLR outer boundary and an important
constraint for models of the BLR. Dust-delay measurements now exist
for $\sim$ a dozen nearby AGN (Koshida et~al.\ 2014; Suganuma
et~al.\ 2006; Minezaki et~al.\ 2004; Oknyanskij \& Horne 2001; Clavel,
Wamsteker, \& Glass 1989; references therein). Most recently,
inter-band continuum delays at shorter wavelengths have been used to
probe the disk radial temperature profile, and if the black hole mass
is known, the mass accretion rate through the disk (Collier 2001;
Collier \& Peterson 2001; Sergeev et~al.\ 2005; Cackett et~al.\ 2007;
Troyer et~al.\ 2016; Edelson et~al. 2015, 2017, 2018; Fausnaugh
et~al.\ 2016; Starkey et~al.\ 2017; Cackett et~al.\ 2018, McHardy
et~al. 2018). Thus RM can be used to study the dominant components
occupying the central regions of AGN on small (disk), intermediate
(BLR), and large (dusty torus) scales, and the relationship(s) between
them.

\subsection{The AGN STORM Campaign}

NGC~5548, a nearby Seyfert~1.5, has a well-documented history of short
timescale large amplitude continuum and broad emission line
variability, and is thus an ideal target for AGN variability campaigns
(e.g., Wamsteker et~al.\ 1990; Clavel et~al.\ 1991; Korista
et~al.\ 1995; Peterson et~al.\ 2002, and references therein).  From
2014 February--July, NGC~5548 was the subject of the most ambitious
spectroscopic reverberation mapping campaign to date, the AGN Space
Telescope and Optical Reverberation Mapping project (AGN STORM,
Peterson PI), with daily monitoring with \textit{HST}/COS over 179
days, yielding 171 usable epochs.  The \textit{HST}/COS program was
amply supported by an intensive \textit{Swift}/XRT/UVOT photometric
monitoring campaign (Edelson, PI) and ground-based optical photometry
and spectroscopy (De Rosa et~al.\ 2015; Edelson et~al.\ 2015;
Fausnaugh et~al.\ 2016, Pei et~al.\ 2017, for details). In addition, 4
\textit{Chandra} X-ray spectra were obtained, over the time period
spanned by the campaign (Mathur et~al.\ 2017).

Key results from AGN STORM include:

(i) identification of a stratified BLR with mean broad emission-line
delays spanning 2.5--7~days with respect to the $\lambda$1157\AA\/
continuum, depending on the emission line (De Rosa et~al.\ 2015).
When placed in context of previous intensive monitoring campaigns of
this source (e.g., Clavel et~al.\ 1991; Peterson et~al.\ 1991, 2002;
Korista et~al.\ 1995; Denney et~al.\ 2009), the measured delays and
variability amplitudes (i.e., responsivities) for the strong UV and
optical broad emission lines for NGC~5548 during 2014 are
significantly smaller, even though the continuum luminosity during
2014 is comparable to that in previous campaigns;

(ii) line of sight velocity versus mean delays indicative of a
radially decreasing velocity field (De Rosa et~al.\ 2015);

(iii) a robust detection of X-ray--UV--optical continuum interband
delays (Edelson et~al.\ 2015; Fausnaugh et~al.\ 2016, Starkey
et~al.\ 2017);

(iv) the $U$ band lag is a factor of $>$ 2 longer than the surrounding
continuum bands ($UVW1$ and $B$) and in excess of the delay predicted
from fitting a simple disk-reprocessing model to the data. This is a
clear indication that diffuse continuum emission (hereafter, DC) from
the BLR (Korista \& Goad 2001) affects the continuum interband lags
(e.g., Edelson et~al.\ 2018);

(v) the appearance of anomalous behaviour in the broad emission lines
starting mid-way through the campaign and lasting $\approx$60~days
(Goad et~al.\ 2016; Pei et~al.\ 2017), as described below. Almost
certainly related to this phenomenon is the sudden increase in the
soft X-ray emission prior to the onset of the anomaly followed by a
more gradual decline (Mathur et~al.\ 2017).

Detailed studies of the time and wavelength-dependent continuum
behaviour have the potential to reveal the nature of the central
engines of AGN: the accretion structure, the means of energy
propagation within the accretion structure, and the origin of the
ubiquitous luminous flux variability in AGN. Significantly, the
measured inter-band continuum variations during the AGN STORM campaign
were characterised by increased delays and reduced amplitude
variations at longer wavelengths and, though qualitatively consistent
with a simple thin-disk reprocessing scenario, imply a disk size in
NGC~5548 which is a factor few too large for its luminosity (Edelson
et~al.\ 2015; Fausnaugh et~al.\ 2016; Starkey et~al.\ 2017), similar
to that found from disk reverberation mapping studies of other nearby
AGN (Cackett et~al.\ 2007, 2018; Troyer et~al.\ 2016). Quasar
micro-lensing studies also point to larger than expected disk sizes
(Poindexter et~al.\ 2008; Morgan et~al.\ 2010; Mosquera
et~al.\ 2013). While alternate explanations abound (e.g., Gardner \&
Done 2017; Dexter \& Agol 2011), interpretation of the measured delay
signatures first requires careful accounting of other known variable
contributions to the observed continuum, such as from the BLR (see
e.g., Korista \& Goad 2001; hereafter KG01).

\begin{figure*}
\begin{center}
\includegraphics[angle=270,scale=0.61]{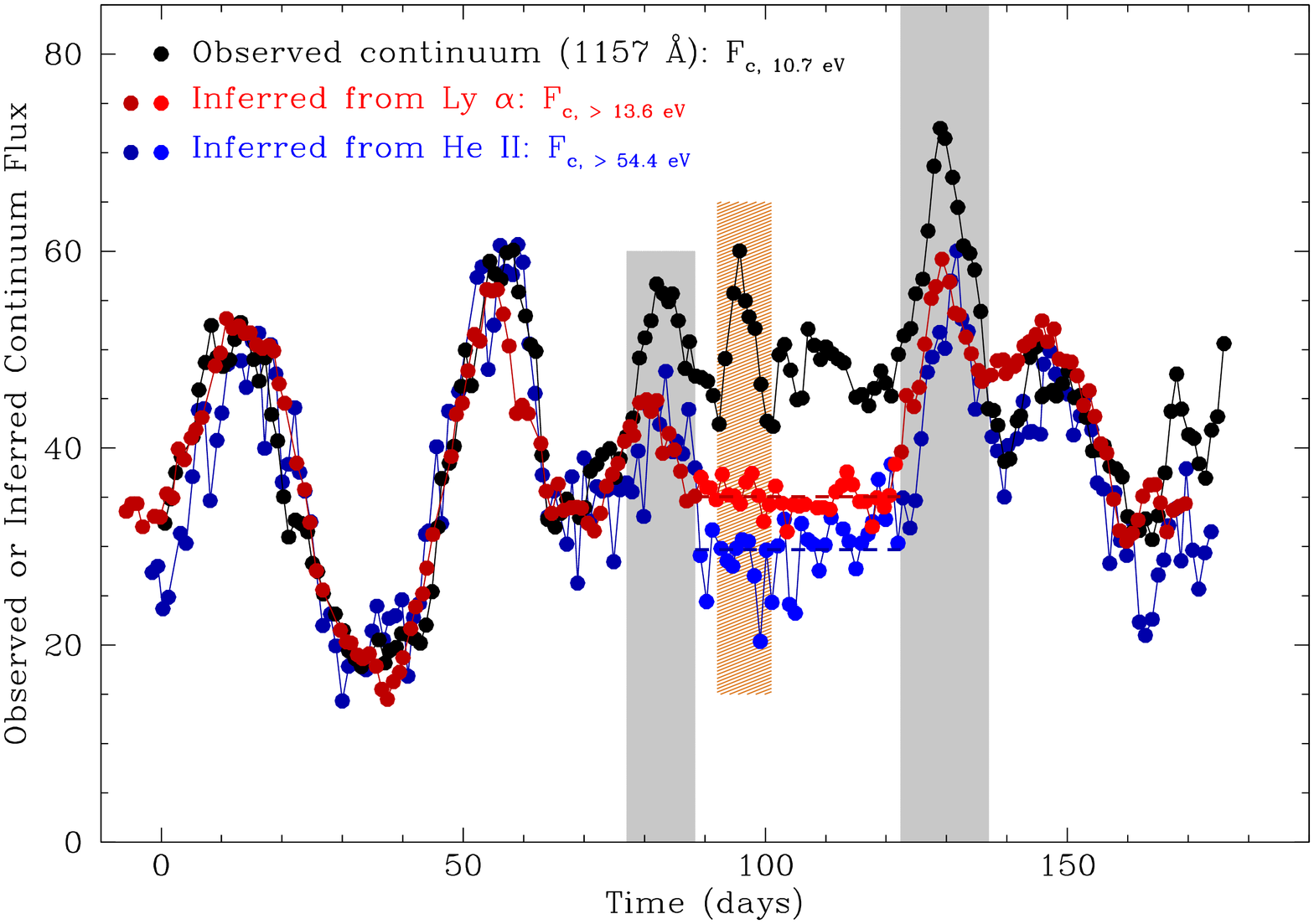}
\caption{A comparison between the \textit{HST}/COS $\lambda$1157\AA\/
  continuum band (black points) and our inferred proxy continuum
  responsible for driving the Ly$\alpha$ (red points) and He~{\sc ii}
  $\lambda$1640\AA\/ (blue points) emission-line variations. Grey
  shaded regions mark the time of approximate onset and exit of the
  anomaly. The orange shaded region, centred near day 95, indicates
  the most significant continuum event (at $\lambda$1157\AA\/) that
  occurred during the period of anomalous emission-line behaviour. The
  dashed horizontal lines indicate the inferred average values found
  during the anomalous period, colour-coded as above.}
\label{fig1}
\end{center}
\end{figure*}

Of particular interest here is the appearance mid-way through the
campaign of anomalous behaviour in the UV/optical broad emission lines
relative to the observed far-UV continuum (Figure~\ref{fig1}, \S2.1;
see also Goad et~al.\ 2016; Pei et~al.\ 2017).  This behaviour is
observed in all of the strong UV and optical broad emission lines and
is characterised by a significant drop in emission line flux and an
absence of response to continuum variations lasting $\approx$40 days,
bracketed by two shorter duration (10--15~days) ``transition'' periods
which trace the departure from ``normal'' emission line behaviour
into, and then out of, the anomaly. These transition periods are
characterised by a reduced emission line flux and a weaker, though
non-zero, response to continuum variations.

The appearance of the anomalous emission-line behaviour mid-way
through the 2014 campaign is the first reported incidence of a
breakdown in two of the fundamental assumptions that underpin broad
emission-line reverberation mapping studies, specifically:

\begin{enumerate}
\item[]{(1) There exists a simple causal relationship between the
  observed continuum and broad emission-line flux variations, and}
\item[]{(2) The observed continuum variations are a smoothed and
  scaled version (and are thus a suitable proxy) for the unobserved
  driving continuum variations at ionising energies.}
\end{enumerate}

\noindent This new and unexpected behaviour may provide new insights into the generation and transport of energy in the vicinity of the black hole.

\subsection{The anomalous behaviour as a diagnostic tool}

Goad et~al.\ (2016) proposed two possible scenarios for the anomalous
broad emission line behaviour: (i) temporary obscuration by a veil of
material lying between the ionising continuum source and the inner BLR
that absorbs some fraction of the incident ionising continuum, or (ii)
an intrinsic change in the strength, shape and variability behaviour
of the ionising continuum. Since the anomaly is seen in all of the
broad emission lines, though with differing flux deficits, any
obscuring veil must be located at small BLR radii. The broad emission
line flux deficits during the anomalous period are large,
$\approx$25\% for the higher ionisation lines, requiring a large solid
angle for the obscuring material. However, while an obscuring veil
could reproduce the observed drop in emission line strengths, it would
not readily explain the absence of response of the broad emission
lines to the modest, yet significant, UV--optical continuum variations
that were present during the anomalous period, for example, the large
continuum event starting at HJD~2,456,785 (day 94, refer to
Figure~\ref{fig1} and \S2.1).  Consequently, Goad et~al.\ (2016)
favoured an intrinsic change in the strength and shape of the ionising
continuum as the more likely explanation (see also Mathur
et~al.\ 2017).  Whatever is the origin, it is evident that during the
anomalous period much of \textit{the ionising continuum incident upon
  the BLR de-cohered in behaviour from the far-UV continuum along our
  line of sight, both in average flux level and in variability} (Goad
et~al.\ 2016).  This is the subject of future work.  Here, we pursue
the use of the anomalous behaviour of the broad emission lines to
illuminate two important aspects of the variable continuum.

First, Goad et~al.\ (2016) used the observed broad emission line
behaviour pre- and post- anomaly as a proxy for estimating the
``expected'' emission line behaviour based on the observed far--UV
($\lambda$1157\AA\/) continuum variations as though the anomaly had
not occurred. From this they were able to estimate the flux deficit in
each of the broad emission lines during the anomalous period.  Here we
adopt an alternative approach, and use key diagnostic emission lines
to trace the changes in the incident SED at ionising energies that
\textit{must have occurred} in order to produce the emission line
behaviour that we observe. The importance of this difference in
approach cannot be understated. Revealing the underlying behaviour of
the ionising continuum is a key ingredient for recovering the BLR
transfer function (e.g., Horne, Welsh, \& Peterson 1991; Welsh \&
Horne 1991), and for constructing dynamical models of the BLR
(Pancoast et~al.\ 2012, 2014a,b; Barth et~al.\ 2011, 2013)\footnote{A
  key assumption underpinning this work is that the time-variable
  behaviour of the shortest wavelength continuum band at
  $\lambda$1157\AA\/ ($\lambda$1138\AA\/ in the rest-frame of the
  source), is a suitable proxy for the continuum at ionising energies
  during times \textit{outside of the anomaly}. While this band lies
  closest to the driving continuum and is the least contaminated by
  stellar light from the host galaxy, it is likely a temporally
  smoothed, reduced-in-amplitude, and slightly delayed version of the
  actual driving continuum.}, and is thus a necessary step toward
achieving one of the principal goals of the AGN STORM
campaign. Second, we search for the signature of the anomalous
behaviour within the measured \textit{continuum bands} to identify the
presence of continuum light emanating from the BLR.  These are
described in turn in \S2 and \S3, below.

\section{Inferring a proxy for the ionising continuum light curve}

To infer a proxy for the ionising continuum's light curve, and in so
doing track changes to its spectral energy distribution (hereafter,
SED) incident upon the BLR, we analyse the light curves of two key
diagnostic emission lines, Ly$\alpha$ and He~{\sc ii}
$\lambda$1640\AA\/.  Ly$\alpha$, a resonance line and among the
strongest UV emission lines, is sensitive to the number of ionising
photons at energies above the Lyman limit (E$>$13.6~eV). He~{\sc ii}
$\lambda$1640\AA\/, on the other hand, is a high-excitation
recombination line and is sensitive to the flux of photons with
energies above 54.4~eV. He~{\sc ii}, though far weaker than
Ly$\alpha$, has a small lag ($\approx$2.5--3~days) relative to the
shortest accessible UV continuum band ($\lambda$1157\AA\/) and
consequently is expected to suffer little geometric
dilution\footnote{Photoionisation model calculations and observations
  indicate that the broad emission-lines originate over a broad range
  in physical conditions and distances from the central source of
  ionising photons.  That being the case, the observed emission-line
  light curve will in general be \textit{reduced in amplitude} and
  delayed in time, depending on details of the mapping between
  source--cloud distance and the observer time-delay, as well as the
  characteristic time scale of the continuum variations relative to
  the relevant light travel times for a particular emission line
  (e.g., Goad \& Korista 2014). Together, these effects are referred
  to as \textit{geometric dilution} of the emission line response.} in
its response. Other than a small shift in delay, He~{\sc ii} is
therefore expected to faithfully track the ionising photon flux at
higher energies.

Assuming that the $\lambda$1157\AA\/ continuum band faithfully tracks
the continuum at ionising energies\footnote{The inferred driving
  continuum light-curve determined by Starkey et~al.\ (2018) for this
  campaign, does not resemble the soft or hard band X-ray
  light-curves, but is instead a good proxy for the 1157\AA\/
  light-curve, further justifying this assumption.}, and that the
emission-line variations arise from a narrow range in delays, we can
infer a suitable proxy for the driving continuum as seen by the broad
emission line region by simply shifting in time and scaling in
amplitude, the observed emission line light curve to match the
observed UV continuum variations outside of the anomalous period (see
e.g., Goad et~al.\ 2016, their Figure~1c).  The required scale factor
for each emission-line, after normalising to their mean values, is
simply the inverse of the responsivity, the power-law index relating
the observed continuum and broad emission line fluxes, as reported by
Goad et~al.\ (2016), their Table~2\footnote{Alternatively, the
  continuum and emission line light curves can be scaled to one
  another using the ratio of their root mean square variations. These
  are mathematically equivalent for small variations about some
  average value (Krolik et~al.\ 1991).}. The smaller than expected
delays and emission-line responsivities, and consequently large scale
factors, likely indicate \textit{a significant contribution to the
  line emission from a large reservoir of gas which is largely
  unresponsive} to the uncharacteristically short timescale (relative
to previous campaigns) ionising continuum variations. That is, the BLR
in NGC~5548 is much larger than would be inferred from a simple
cross-correlation analysis of the UV continuum and emission-line light
curves. This conclusion is supported by results from reverberation
mapping analyses of this source, which indicate that the BLR in
NGC~5548 is spatially extended (Horne et~al.\ 2019, AGN STORM
paper~{\sc ix}, in prep). The recovered 1-d response functions
$\Psi(\tau)$ span delays of 0 to 50 days, and though the bulk of the
response is on short timescales, there is significant response for
many lines, but particularly for H$\beta$, on timescales as large as
$\sim$40 days. This is to be compared to a CCF centroid of $\sim 6$
days for H$\beta$ relative to the 1157\AA\, continuum band for the
full campaign (Pei et~al. 2017).

In Figure~\ref{fig1} we show the \textit{HST}/COS $\lambda$1157\AA\/
UV continuum light curve (black points) together with the inferred
proxy continua responsible for driving the observed emission-line
variations in Ly$\alpha$ $\lambda$1216 (red points) and He~{\sc ii}
$\lambda$1640\AA\, (blue points).  Delays and scale factors for each
line are here determined for days occurring before the onset of the
anomaly (i.e., before HJD 2456,766.1094, $\approx$75 days after the
start of the \textit{HST}/COS campaign). Remarkably, all that is
required to match well the observed continuum variations outside of
the anomalous period is a simple scaling in amplitude and temporal
shift of the Ly$\alpha$ and He~{\sc ii} broad emission line light
curves. We find little evidence for significant temporal smearing of
the broad emission-line light curves outside of the anomalous period
(compare the C~{\sc iv} emission-line light curve to the one inferred
directly from the $\lambda$1157\AA\/ continuum, Goad et~al.\ 2016,
their Figures 1c,d).  This suggests that the variable part of each
emission-line arises from a surprisingly narrow range in delays ($<$
10~days) relative to the reported lag values, i.e., the
\textit{response} functions must be relatively narrow in time delay.

\begin{figure*}
\begin{center}
\includegraphics[angle=270,scale=0.61]{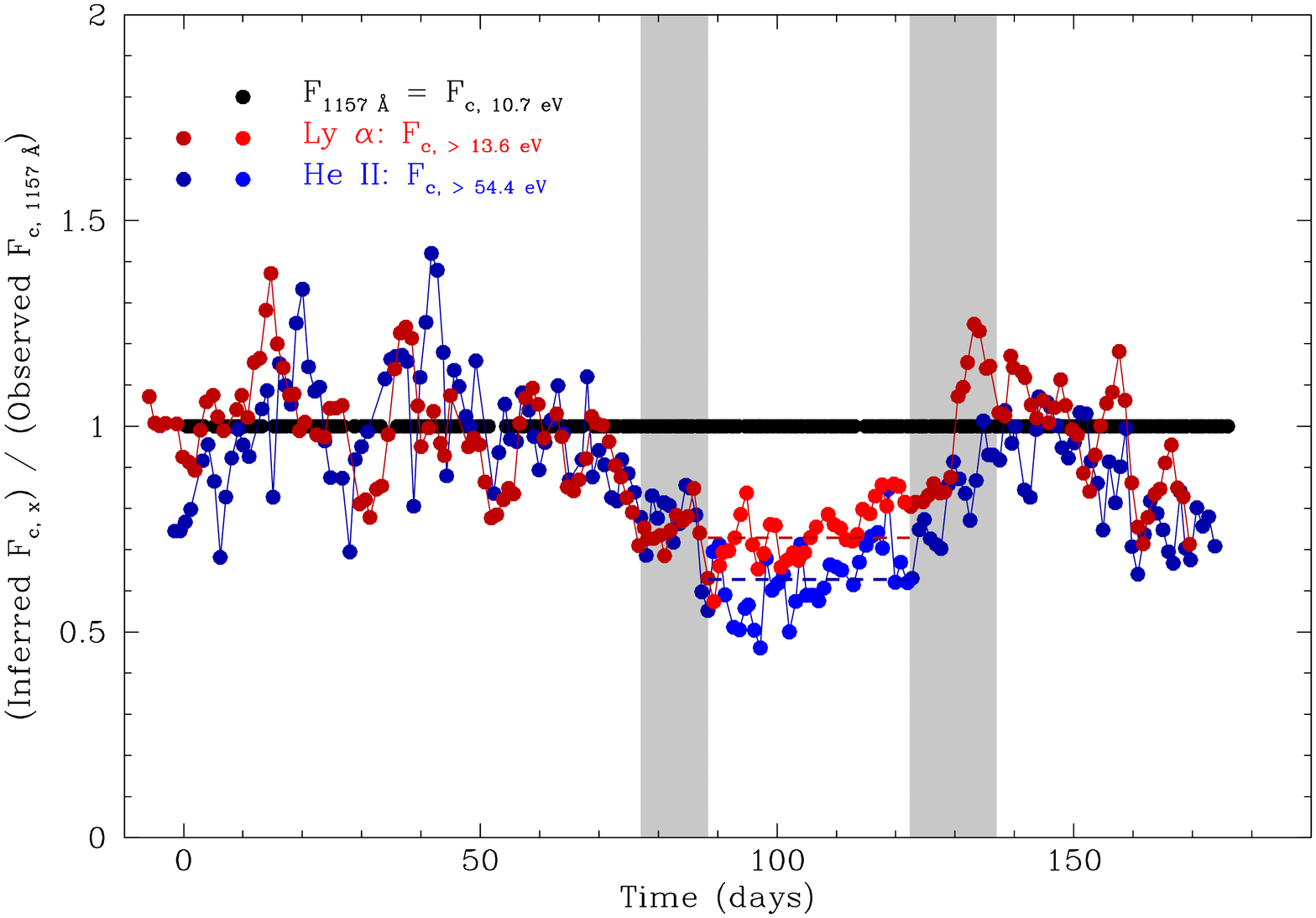}
\caption{The ratio of the inferred proxy driving continuum bands for
  Ly$\alpha$ and He~{\sc ii} $\lambda$1640\AA\/ relative to the
  observed continuum at $\lambda$1157\AA\/. The inferred He~{\sc ii}
  continuum shows a larger decline relative to $\lambda$1157\AA, than
  the inferred Ly$\alpha$ continuum. The dashed horizontal lines
  indicate the inferred average values for this ratio found during the
  anomalous period, colour-coded as above.}
\label{fig2}
\end{center}
\end{figure*}

Figure~\ref{fig1} next illustrates that the most striking difference
between the observed $\lambda$1157\AA\/ continuum light curve and the
continuum that drives the broad emission lines, as inferred from the
behaviour of the Ly$\alpha$ and He~{\sc ii} emission-lines, is the
large deficit in continuum flux: $\approx$30\% for Ly$\alpha$,
$\approx$40\% for He~{\sc ii}, and $\approx$20\% for H$\beta$ (not
shown) during the time of the anomaly.  In Figure~\ref{fig2} we show
the ratio of the inferred proxy driving continua for Ly$\alpha$ (red
points) and He~{\sc ii} (blue points) relative to the observed
$\lambda$1157\AA\/ continuum band.  This indicates a significantly
larger reduction in the continuum flux responsible for driving
variations in the broad He~{\sc ii} emission line relative to the
continuum band at $\lambda$1157\AA\/ than there is for that driving
Ly$\alpha$. Thus during the anomaly the ionising continuum as incident
upon the BLR became \textit{weaker and softer} relative to the
continuum at $\lambda$1157\AA\/. We note here that this analysis is
based on integrated broad line fluxes in Ly$\alpha$ and He~{\sc ii}.
The actual reductions in the 1 and 4 Rydberg continua photon fluxes
incident upon the BLR are therefore likely somewhat larger than the
values quoted above, which are geometrically diluted by effects
associated with light travel time.  As a matter of fact, this same
analysis technique, when applied to individual radial velocity bins,
finds larger flux deficits in the line wings of C~{\sc iv} than in the
line core, consistent with this inference. We will explore these
aspects in future work.

Finally, also illustrated in Figure~\ref{fig1} are three shaded
regions, two grey shaded regions which identify local peaks in the
$\lambda$1157\AA\/ continuum at the beginning and end of the anomaly,
and an orange shaded region marking the largest $\lambda$1157\AA\/
continuum event that was observed during the anomalous period.  The
grey shaded regions can be identified with the two transition periods
reported by Goad et~al.\ (2016) marking the transitions in to and out
of the anomaly. During the transition intervals, the inferred
Ly$\alpha$ and He~{\sc ii} ionising continuum flux track the continuum
at $\lambda$1157\AA\/, but with a \textit{reduced flux and smaller
  variability amplitude}. Importantly, there is no equivalent feature
in the inferred proxy continua (i.e., from Ly$\alpha$ or He~{\sc ii})
to the large observed continuum event in the $\lambda$1157\AA\/
continuum band indicated by the orange shaded region. As noted by Goad
et~al.\ (2016), and given the apparent small spread in delays for the
broad emission lines referred to previously, we should have been able
to measure significant variations in the broad emission-lines related
to this feature, and yet we do not.  Why the ionising continuum no
longer tracks the $\lambda$1157\AA\/ continuum during the anomalous
period is unclear, but presumably is related to the redistribution of
photons in energy.

In the following section, we utilise both of the above characteristic
differences (shown in Figure~\ref{fig1}) pointing to a decohering of
the bulk of the driving ionising continuum incident upon the BLR from
the observed far-UV continuum during the anomalous period, to identify
anomalous behaviour in the UV-optical continuum band measurements.

\section{Anomalous behaviour in the UV--optical continuum bands}

Thus far, the anomaly has been discussed in terms of its effect upon
the observed broad emission-line flux light curves. Whether the
anomaly is present within the UV and optical continuum bands has not
been explored.  However, a feature of photoionisation model
predictions of dense BLR gas is a significant diffuse continuum, in
addition to broad emission-lines (KG01; Lawther et~al.\ 2018). For
example, KG01, constructed a local optimally emitting cloud model
(hereafter LOC, Baldwin et~al.\ 1995) for the broad emission lines in
NGC~5548 designed to match the observed luminosities of the two
strongest UV emission lines, Ly$\alpha$ and C~{\sc iv} (Clavel
et~al.\ 1991; Krolik et~al.\ 1991; Korista et~al.\ 1995). They found
that their model BLR produces (in addition to emission lines) a
significant thermal diffuse continuum component spanning the
\textit{entire UV--optical--near IR continuum}.  This diffuse
continuum component is largely dominated by the H--He recombination
(free--bound) continuum, with smaller contributions from free--free
emission as well as scattering from free electrons and neutral
hydrogen (see Korista \& Ferland 1998). Since the diffuse continuum
originates in the spatially extended BLR, it will display a broad
range in variability amplitude and delay.
While the lag spectra for both the disk and the DC components
generally increase to longer wavelengths, their detailed shapes
differ. The accretion disk lags are expected to go as $\tau \propto
\lambda^{4/3}$ (Cackett et al. 2007), the DC has more of a sawtooth
shape with peaks e.g., near the Balmer and Paschen breaks (KG01;
Lawther et al. 2018). While the differences are better delineated in
data with much higher spectral resolution (e.g. NGC 4593; Cackett et
al. 2018), it is much harder to resolve with broadband photometry such
as used here.

For conditions appropriate to the BLR, a significant diffuse continuum
component is largely unavoidable.  Yet, interpretation of the larger
than expected continuum inter-band delays have instead preferred to
appeal to alternate models for describing the structure of the disk,
including: truncated disks (Narayan 1996), truncated disks with an
additional reprocessing region (Gardner \& Done 2017), inhomogeneous
disks (Dexter \& Agol 2011), patchy disks (Starkey et~al.\ 2017), disk
atmospheres (Hall et~al. 2018), and tearing disks (Nealon
et~al.\ 2015).  The reduced variability and larger than expected
delays measured at all wavelengths, and in particular in the vicinity
of the Balmer continuum, relative to that predicted by the standard
disk model (Edelson et~al.\ 2015, 2017, 2018; Fausnaugh et~al.\ 2016;
Starkey et~al. 2017, McHardy et~al. 2018) we argue is broadly
consistent with what one might expect if the UV--optical continuum
bands are significantly contaminated by a diffuse continuum emitting
from the same gas that emits the broad emission lines (KG01; Lawther
et~al.\ 2018).

In general, the longer wavelength continuum bands will be reduced in
amplitude and smeared in time relative to the shorter wavelength
continuum bands because:
\begin{itemize}
\item the underlying UV-near-IR continuum arises from thermalised
  emission of an accretion disk (or some other accretion geometry)
  with a spatial gradient in gas temperature, such that kT diminishes
  with increasing radius,
\item whatever the mechanism behind the variations in luminous flux
  from the disk, its transfer function is not a delta-function,
\item there may be a significant DC contribution from the BLR.
\end{itemize}

All of the above may be contributing to the observed continuum
variations at some level, and all will act to reduce the continuum
inter-band correlations.

Here we take advantage of the presence of the anomaly to identify the
presence of the diffuse continuum component. If the diffuse continuum
component is important, we should see the anomaly's effect in the
observed continuum bands, and should follow a pattern of behaviour
that is similar to that seen in the emission lines.

In particular, we predict that the diffuse continuum would have
undergone a {\it significant drop in flux\/} during the anomaly in
contrast to the $\lambda$1157\AA\/ continuum
band\footnote{Photoionisation models indicate that the DC contribution
  to the 1157\AA\ continuum band is small in comparison to most longer
  wavelength UV--optical continuum bands. However, we can not exclude
  the possibility that the 1157\AA\ continuum band itself exhibits
  characteristics of anomalous behaviour of the DC.}. Additionally, we
predict that the most significant continuum event observed during the
anomalous period (Figure~\ref{fig1}, orange shaded region) will be
significantly weaker in the longer wavelength DC band light-curves, as
it is for the integrated flux in broad emission lines.  Because the
behaviour of the DC component is controlled by the same driving
continuum inferred for the broad emission-lines, then any significant
presence of the DC will act to suppress the inter-band continuum
correlations during the anomalous period relative to their behaviour
at other times.

To test this expectation we here investigate the relationship between
the continuum at longer wavelengths relative to the shortest
accessible UV band $\lambda$1157\AA\/, a band for which
photoionisation model calculations suggest has only a weak
contribution from the BLR diffuse continuum (KG01), and which is also
least likely to have contributions from the wings of broad emission
lines, when compared to other continuum bands (Kriss
et~al.\ 2019). Specifically, we test for {\it differential
  behaviour\/} between this band and the longer wavelength continuum
bands for time periods spanning the anomaly and otherwise.

\begin{figure*}
\begin{center}
  \includegraphics[angle=0,scale=0.65]{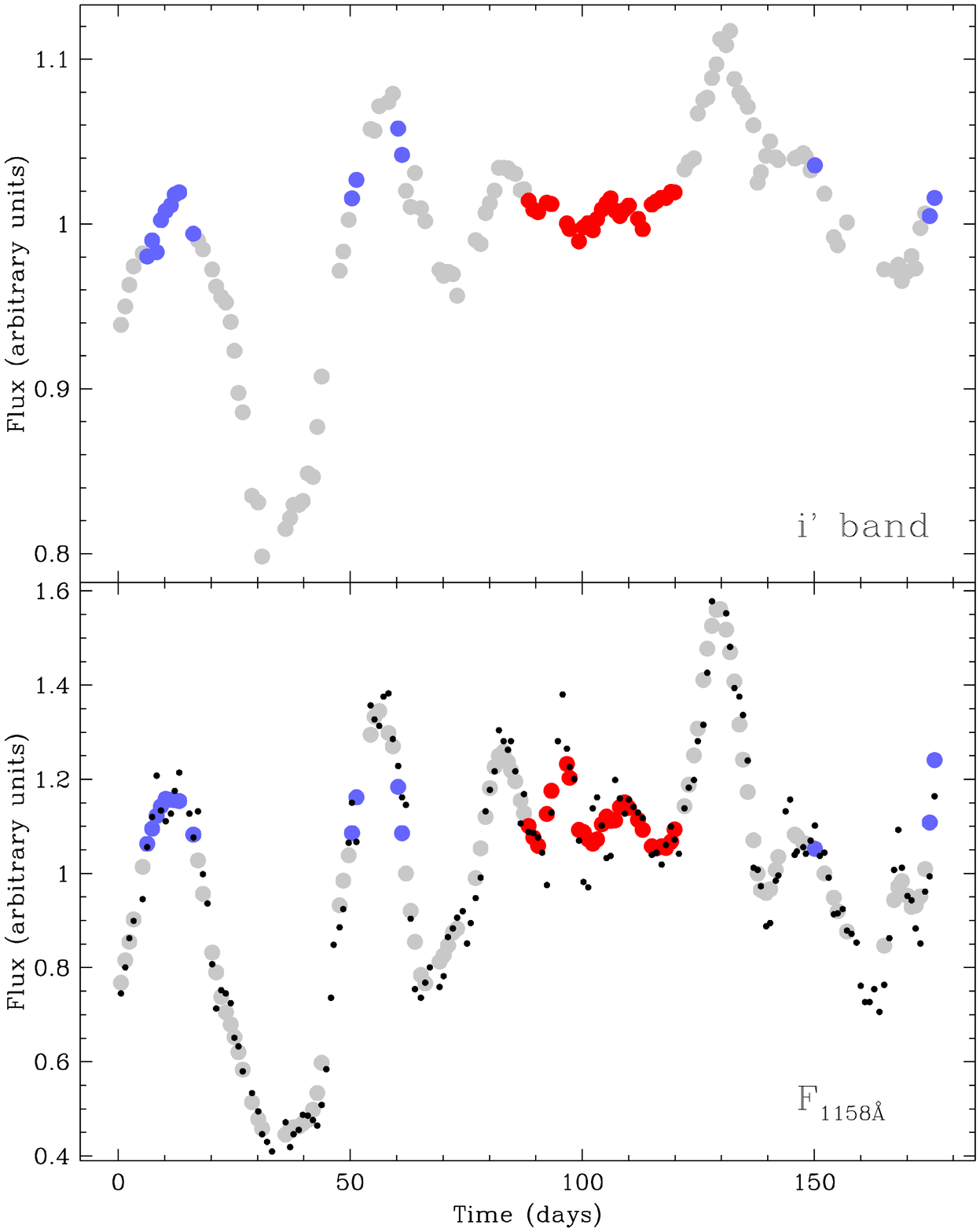}
\vspace{-5mm}
\caption{A comparison between the delay-shifted ground-based
  i$^{\prime}$ continuum light curve (grey points, upper panel), the
  original, unsmoothed $\lambda$1157\AA\, \textit{HST}/COS UV
  continuum (black points, lower panel), and the $\lambda$1157\AA\,
  continuum after convolving with the disk transfer function (grey
  points, lower panel, see text for details). Each light curve has
  been normalised to its mean value. In the lower panel, red and blue
  points indicate portions of the light curve covering similar ranges
  in flux within (red) and outside of (blue) the anomaly. In the upper
  panel red and blue points indicate the corresponding {\it
    i}$^{\prime}$-band points over the same time intervals.}
\label{fig3}
\end{center}
\end{figure*}

The method we employ is illustrated in Figure~\ref{fig3}. First, we
shift each of the longer wavelength continuum bands backwards in time
according to their measured delay with respect to the
$\lambda$1157\AA\/ continuum band. We use the delays reported by
Fausnaugh et~al.\ (2016), and Pei et~al.\ (2017), determined using the
software package {\sc javelin} in order to shift the light
curves. This is mathematically equivalent to assuming the signal is
reprocessed in a face-on ring.  The $\lambda$1157\AA\, continuum band
and its longer wavelength pair are then interpolated onto a common
grid (using simple linear interpolation). Only points which are close
in time ($<0.5$~day difference) and common to both light curves are
plotted (grey points). This avoids extrapolation beyond the ends of
the light curves and ensures that the number of useful pairs of points
is preserved.

Referring to the lower panel of Figure~\ref{fig3}, for each light
curve pair we highlight in red points taken from the middle of the
anomalous period (days 86--123). The blue points in the bottom panel
indicate those days outside of the anomaly for which the
$\lambda$1157\AA\/ continuum band shows approximately the same range
in flux as found during the anomalous period excluding the transition
regions (indicated in red).  This allows us to compare the strength of
the flux--flux correlations inside and outside of the anomalous period
over the same range in flux level as that within the 1157\AA\,
continuum band. Nonetheless, it is the case that the amplitude and
timescale for the variability are smaller on average inside of the
anomaly than outside of the anomaly (e.g., Sun et~al.\ 2018).  Since
we cannot control for this latter effect, and because a reduced
correlation is a natural consequence of reduced variability amplitude
in the presence of noise, the test described here is by no means
definitive, and hence our conclusions are not without bias. We
investigate this further in $\S3.2$.

\begin{figure*}
\begin{turn}{90}
\begin{minipage}{8.5in}
\begin{center}
\includegraphics[width=\textwidth]{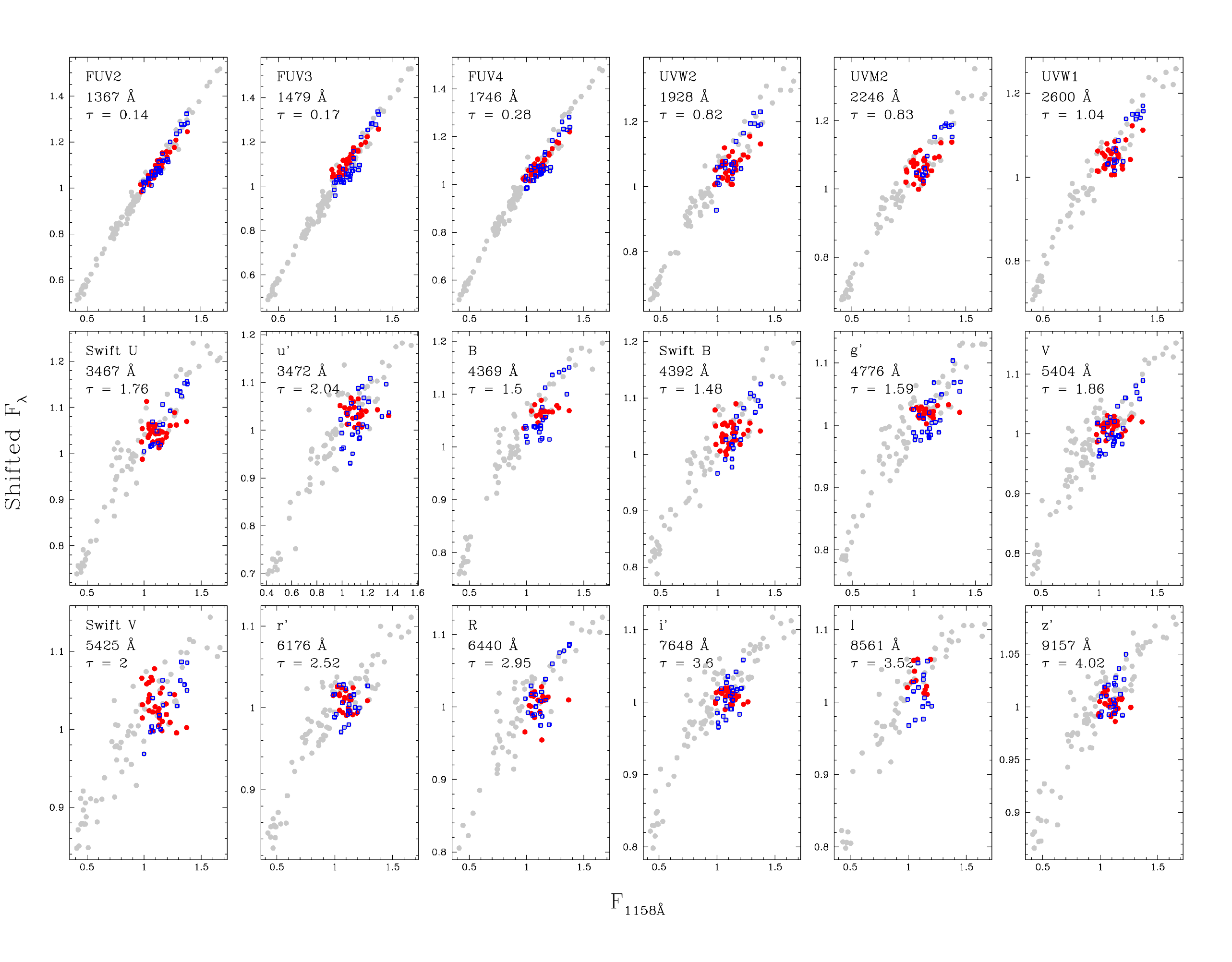}
\vspace{-8mm}
\caption{Flux-flux diagrams for the longer wavelength continuum bands
  relative to the shortest wavelength \textit{HST}/COS
  $\lambda$1157\AA\/ continuum band. The longer wavelength bands have
  first been shifted by their measured delay in days (the reported
  $\tau$ values) relative to 1157\AA\/ continuum (see text for
  details). Flux--flux pairs taken from outside of, and inside of, the
  anomalous period are indicated in blue and red respectively, the
  remainder of the points are marked in grey.}
\label{fig4}
\end{center}
\end{minipage}
\end{turn}
\end{figure*}

\begin{figure*}
\begin{turn}{90}
\begin{minipage}{8.5in}
\begin{center}
\includegraphics[width=\textwidth]{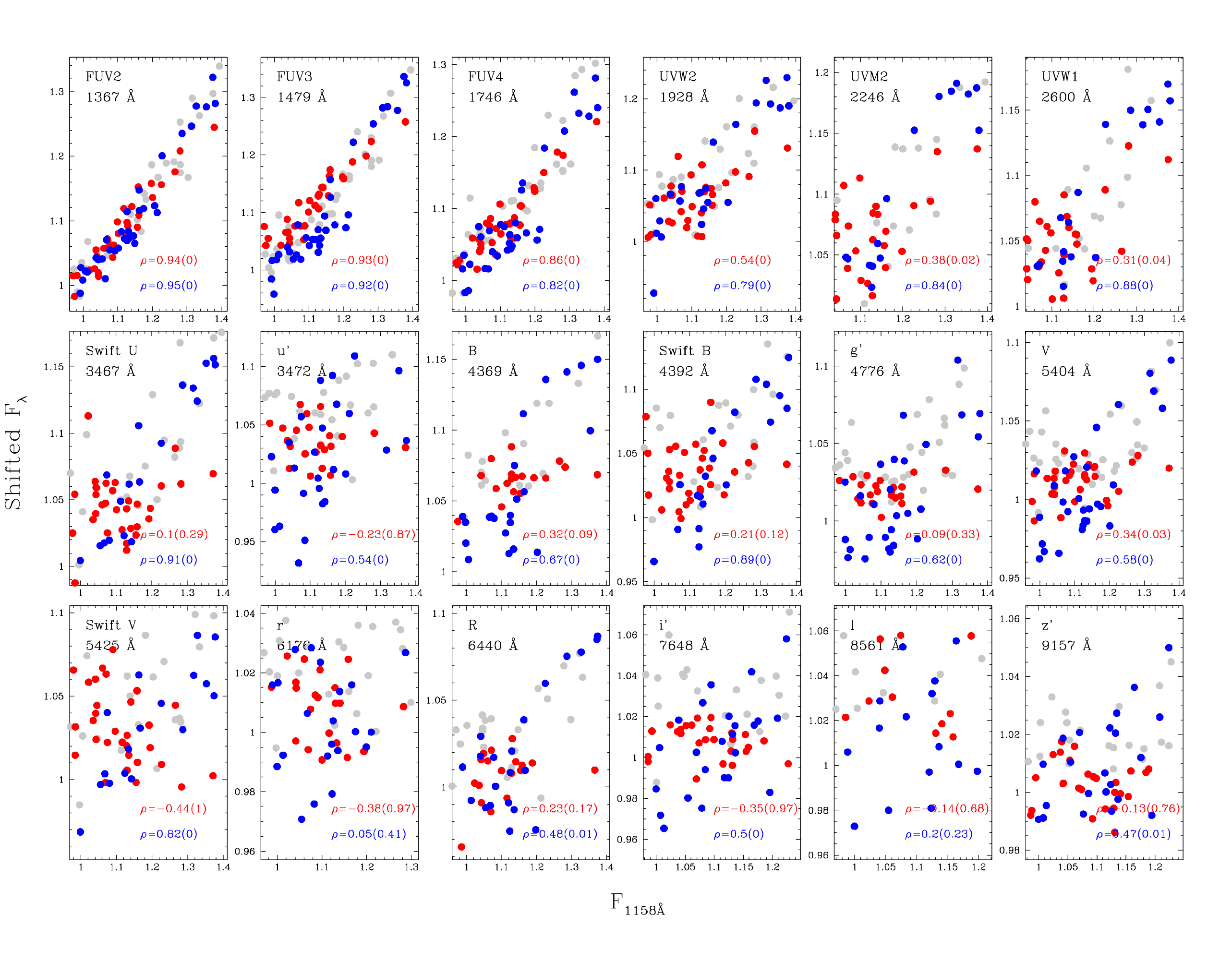}
\vspace{-8mm}
\caption{A zoomed in version of Figure~\ref{fig4}. Reported Spearman's
  rank correlation coefficients between flux-flux pairs, taken over a
  similar range in flux, outside of and inside of the anomalous period
  are indicated in blue and red respectively.}
\label{fig5}
\end{center}
\end{minipage}
\end{turn}
\end{figure*}

In essence the blue points indicate what the ``normal'' behaviour of
the diffuse continuum from the BLR should be for the range in
$\lambda$1157\AA\/ continuum flux present during the anomalous period,
modulo the bias discussed above. The top panel of Figure~\ref{fig3}
identifies the same blue and red points in time for the $i^{\prime}$
band. Light curve pairs in the other continuum bands are then
similarly transposed onto flux--flux diagrams with the same colour
coding (refer to Figures~\ref{fig4} and \ref{fig5}).

Immediately apparent from Figures~\ref{fig4}--\ref{fig5}, is that even
though the blue points are spliced together from five essentially
randomly-selected segments, for the majority of continuum bands, they
are more tightly correlated with each other than the contiguous red
(anomalous period) points. That is, over the observed range in
$\lambda$1157\AA\/ continuum flux present during the anomalous period,
the longer wavelength inter-band continuum variations correlate better
outside of, rather than inside of the anomalous period. We suggest
that the UV (perhaps also including $\lambda$1157\AA\/) and optical
continuum bands therefore must also be affected during the anomaly,
and in a similar fashion to the emission lines.

To quantify the extent to which the longer wavelength continuum bands
decorrelate from the $\lambda$1157\AA\/ continuum band during the
anomalous period we perform a simple Spearman rank correlation between
the flux--flux pairs for the red and blue points. These are indicated
in Figure~\ref{fig5}, alongside zoomed in versions of the flux--flux
correlations\footnote{Spearman rank correlation coefficients are
  invariant to transformations of the form $y \rightarrow y - a$ or $y
  \rightarrow y/a$ and are thus insensitive to the presence of
  non-variable constant components (e.g., the host galaxy
  contribution).}.  The measured correlation coefficients $\rho$ and
p-values (in brackets) are indicated in Figure~\ref{fig5} and
summarised in Table~\ref{tab1} (columns 4, and 5). The correlation
coefficients for the blue points are in the majority of cases larger
than those for the red points, indicating that the variations in the
longer wavelength continuum bands are indeed more weakly correlated
with the $\lambda$1157\AA\/ continuum band during the anomalous
period. Since the red and blue points identified in individual
light-curves are drawn from the same range in flux level and are thus
of similar signal-to-noise, the weaker correlation found between
flux-flux pairs taken from inside of the anomaly must arise due to
intrinsic differences in light-curve behaviour.

While a reduced correlation between the short and longer wavelength
bands arises naturally from a reverberating disk, this cannot explain
the \textit{difference in behaviour} inside and outside of the
anomalous period.  Significant contributions from a diffuse continuum
emanating from BLR clouds would also act to further reduce the
inter-band correlations. However, the observed differential behaviour,
i.e., the reduced inter-band correlations found during the anomaly
(red points) relative to those found outside of the anomaly (blue
points) we suggest arises because the continuum bands contain a
substantial amount of light from the BLR clouds, \textit{which as for
  the emission-lines, underwent a drop in flux and did not see the
  significant continuum event} noted in Figure~\ref{fig1}, centred on
day 95 (orange shaded region).

Finally, we note that the contribution of non-variable sources (e.g.,
the host galaxy contribution, and reprocessed continuum from narrow
line region gas) to the flux in the continuum bands is also likely
wavelength dependent. However, our simple ranking scheme for measuring
the correlation strength between the continuum bands is insensitive to
these non-variable components.

\begin{table*}
  \begin{center}\caption{Mean delays and Spearman rank correlation coefficients for the measured UV--optical--near IR continuum bands\label{tab1}}
\begin{tabular}{lrrrrrr}
\hline \hline Bandpass & pivot $\lambda$ & delay$^{\dagger}$ &
$\rho$ & $\rho$ & $\rho$ (sm)$^{\ddagger}$ & $\rho$ (sm) \\ & (\AA\,) & (days) &
(red) & (blue) & (red) & (blue)\\ \hline HST/COS & 1367 & 0.14 &
0.94 & 0.95 & 0.94 & 0.96 \\ HST/COS & 1479 & 0.17 & 0.93 & 0.92 &
0.94 & 0.93 \\ HST/COS & 1746 & 0.28 & 0.86 & 0.82 & 0.87 & 0.82
\\ Swift UVW2 & 1928 & 0.82 & 0.54 & 0.79 & 0.54 & 0.77 \\ Swift UVM2
& 2246 & 0.83 & 0.38 & 0.84 & 0.37 & 0.73 \\ Swift UVW1 & 2600 & 1.04
& 0.31 & 0.88 & 0.29 & 0.82 \\ Swift U & 3467 & 1.76 & 0.10 & 0.91 &
0.12 & 0.83 \\ u$^{\prime}$ & 3472 & 2.04 & $-$0.23 & 0.54 & $-$0.06 &
0.51 \\ B & 4369 & 1.50 & 0.32 & 0.67 & 0.38 & 0.39 \\ Swift B & 4392
& 1.48 & 0.21 & 0.89 & 0.23 & 0.89\\ g$^{\prime}$ & 4776 & 1.59 & 0.09
& 0.62 & 0.23 & 0.18 \\ V & 5404 & 1.86 & 0.34 & 0.58 & 0.44 & 0.28
\\ Swift V & 5425 & 2.00 & $-$0.44 & 0.82 & $-$0.41 & 0.62
\\ r$^{\prime}$ & 6176 & 2.52 & $-$0.38 & 0.05 & $-$0.17 & $-$0.03
\\ R & 6440 & 2.95 & 0.23 & 0.48 & 0.37 & 0.39 \\ i$^{\prime}$ & 7468
& 3.60 & $-$0.35 & 0.50 & $-$0.18 & 0.32 \\ I & 8561 & 3.52 & $-$0.14
& 0.20 & $-$0.65 & 0.37 \\ z$^{\prime}$ & 9157 & 4.02 & $-$0.13 & 0.47
& $-$0.06 & 0.24 \\ \hline
\end{tabular}
  \end{center}
\noindent  $^{\dagger}$Measured delays are from Fausnaugh et~al.\ (2016), and Pei et~al. (2017) and are measured relative to the \textit{HST}/COS $\lambda$1157\AA\/ \flushleft continuum band.

\noindent  $^\ddagger$(sm) indicates that the $\lambda$1157\AA\/ light-curve has first been convolved with a disk transfer function (see text for details).
\end{table*}

\begin{figure*}
\begin{turn}{90}
\begin{minipage}{8.5in}
\begin{center}
\includegraphics[width=\textwidth]{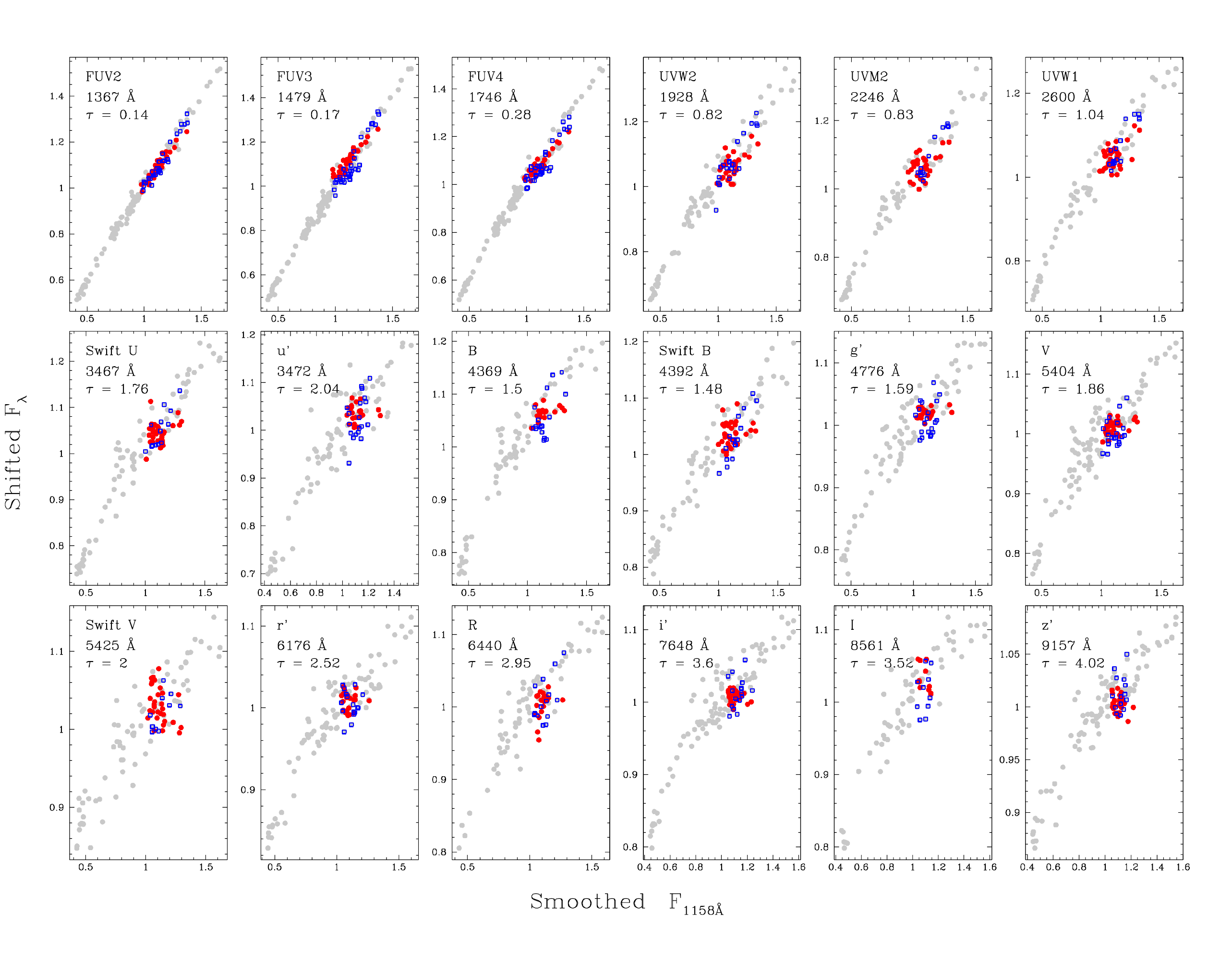}
\vspace{-8mm}
\caption{As for Figure~\ref{fig4}, after first convolving the
  1157\AA\/ continuum band with a model transfer function. For the
  transfer function, we use a thin ring inclined at an angle of 36
  degrees with respect to the observers line of sight and with a mean
  delay representative of the lag between the two continuum bands (see
  text for details).}
\label{fig6}
\end{center}
\end{minipage}
\end{turn}
\end{figure*}

\begin{figure*}
\begin{turn}{90}
\begin{minipage}{8.5in}
\begin{center}
  \includegraphics[width=\textwidth]{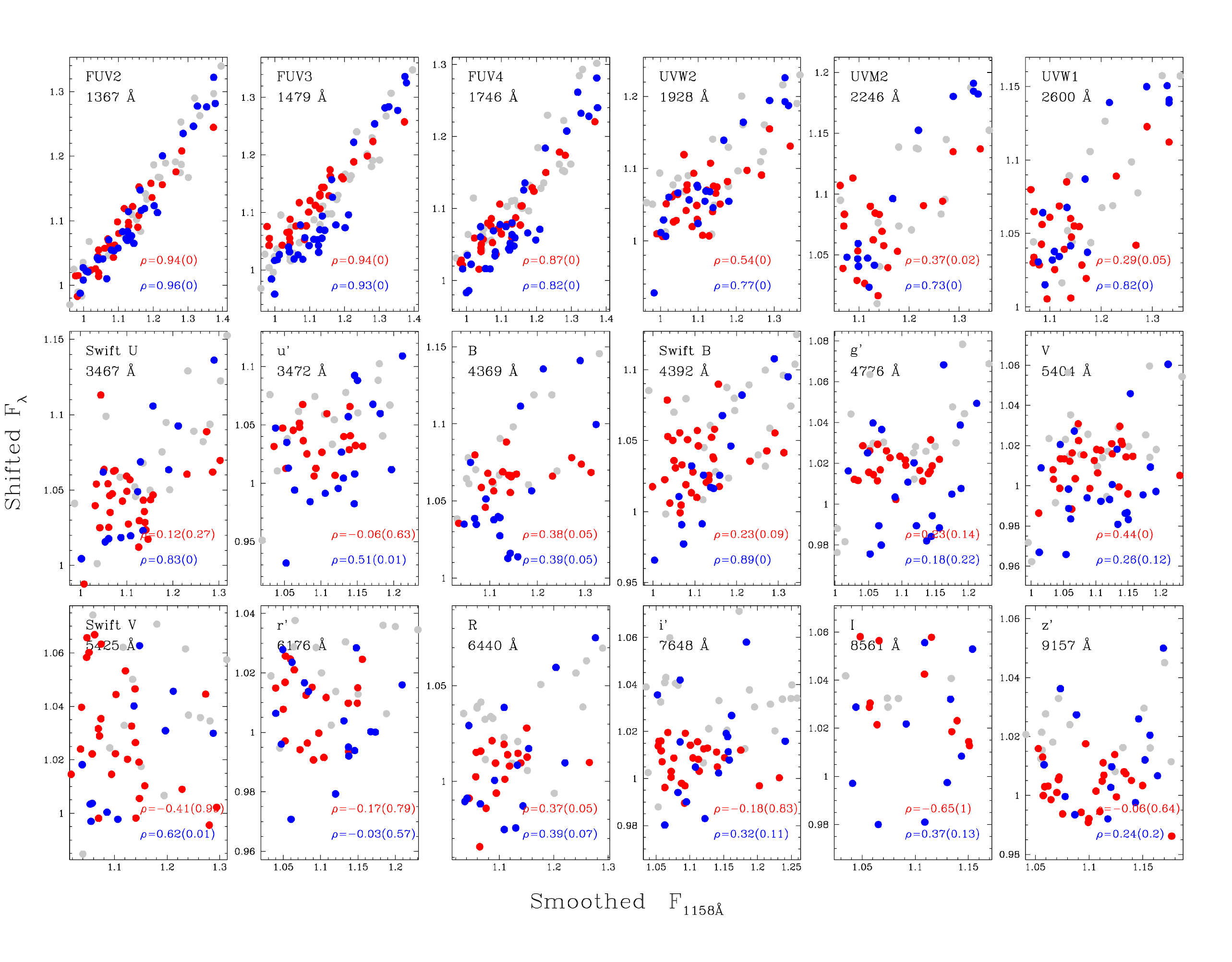}
\vspace{-8mm}
\caption{A zoomed in version of Figure~\ref{fig6}. Spearman's rank
  correlation coefficients and P-values between flux--flux pairs, and
  spanning the same range in flux, outside of and inside of the
  anomaly are indicated in blue and red respectively.}
\label{fig7}
\end{center}
\end{minipage}
\end{turn}
\end{figure*}

\subsection{Disk transfer functions}

Shifting the longer wavelength continuum light curves in time, using a
one-number estimate for the ``size'' of the continuum emitting region
(relative to that emitting $\lambda$1157\AA\/) is a rather simplistic
approach.  In the standard disk model (Shakura \& Sunyaev 1973) each
continuum band arises from a range in disk radii such that the longer
wavelength continuum bands are not only delayed relative to those at
shorter wavelengths, but their variations are also smaller in
amplitude and smoothed in time, reducing the correlation between the
bands (Collier 2001; Cackett et~al.\ 2007, 2018; Edelson et~al.\ 2015;
Fausnaugh et~al.\ 2016; Starkey et~al.\ 2017).  We test the
significance of this ``reduced correlation'' by convolving the
$\lambda$1157\AA\/ continuum light-curve with a model disk transfer
function representing the response of a thin ring of material, with a
size that corresponds to the measured lag between the longer
wavelength continuum band and $\lambda$1157\AA\/ and viewed at and
angle of 36 degrees relative to our line of sight to the disk. This
inclination is equal to the best-fit value found for the disk
orientation in NGC~5548 by Starkey et~al.\ (2017), $36\pm 10$ degrees,
when fitting for the disk radial temperature profile, inclination $i$
and normalisation, similar to the inclination angle derived for the
broad emission-line region in NGC~5548, $38.8^{+12.1}_{-11.4}$ degrees
(Pancoast et~al.\ 2014), and in broad agreement with the inclination
required to recover the virial factor \textit{f} for this source (Goad
et~al.\ 2012), and the inclination inferred for this source from the
recovered broad emission-line velocity delay maps (Horne et~al. 2019,
AGN STORM paper {\sc ix}, in prep.). This smoothed version of the
$\lambda$1157\AA\/ continuum light-curve is then correlated against
the longer wavelength continuum band following the procedure outlined
above.

The results of this process are shown in
Figures~\ref{fig6}--\ref{fig7}, and summarised in Table~\ref{tab1}
(columns 6 and 7). Convolving with a disk transfer function acts to
tighten the relationship between the $\lambda$1157\AA\, and longer
wavelength continuum bands. We also refer the reader back to
Figure~\ref{fig3} for a direct comparison between the
$\lambda$1157\AA\/ continuum light curve and that of one at longer
wavelength. However, while the correlation coefficients are slightly
modified, our main result still stands. Red points taken during the
anomalous period are consistently less correlated with the
$\lambda$1157\AA\/ continuum band than are the blue points. We note
that Sun et.~al.\ 2018 report significant differences between the
delay and response amplitude between the \textit{Swift} X-ray and far
UV \textit{HST} band ($\lambda$1367\AA\/), indicating that the delays
between these two bands are longer and more tightly correlated
(coherent) in the normal state (outside of the anomaly) than in the
anomalous state, further justifying our claim.

\subsection{Additional tests}
We perform one additional test, assuming that the longer wavelength
continuum light-curves are scaled, smoothed and delayed versions of
the continuum bands at shorter wavelengths. We use the shortest
wavelength continuum band from \textit{HST} ($\lambda$1157\AA\/) as a
reference, as photoionisation model calculations suggest that this
band is likely least contaminated by diffuse continuum emission from
BLR gas (KG01; Lawther et~al.\ 2018). We then fix the delay of the
longer wavelength bands relative to this reference band using values
reported in Fausnaugh et~al.\ (2016), and then scale the \textit{HST}
band according to the relation

\begin{equation*}
\log F_{\lambda}({\rm long}) = \alpha \log F_{\lambda}({\rm short}) + K \; .
\end{equation*}

\begin{figure*}
  \begin{center}
\includegraphics[angle=0,scale=0.65]{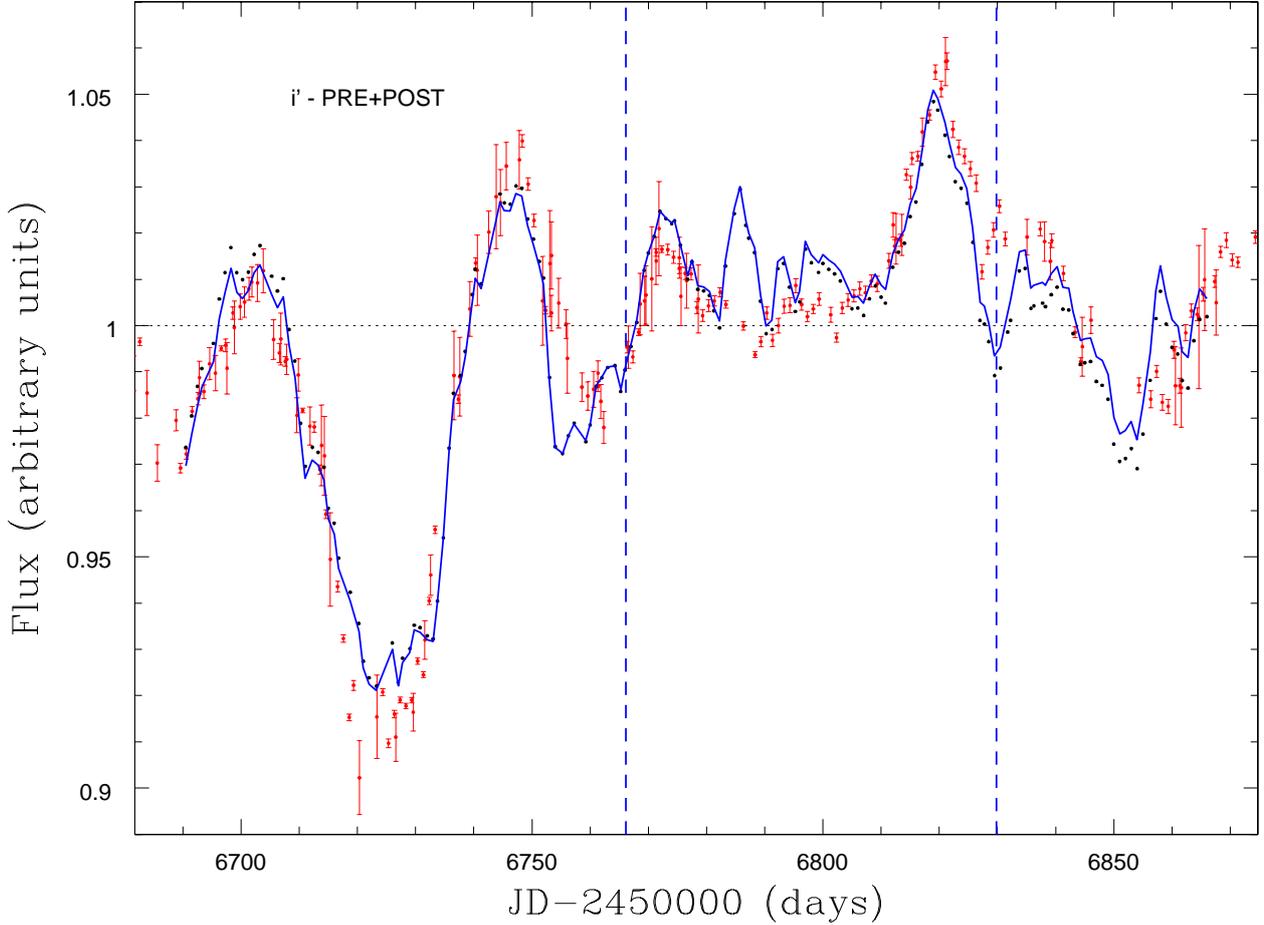}
\vspace{-5mm}
\caption{Ground-based i$^{\prime}$ broadband continuum light curve
  (red points), matched to a shifted, scaled and smoothed version of
  the \textit{HST} 1157\AA\/ continuum band light curve (black
  points). The solid blue-line indicates the best-fit model obtained
  when an additional linear background component is included (see text
  for details). The dashed blue vertical lines indicate the time
  period spanning the anomalous behaviour of the broad emission-lines
  reported by Goad et~al.\ (2016) (epochs JD
  2456766.1--2456829.8). This region is excluded from the fit
  statistic. Lower-panel - the time variation of the $\chi^{2}$
  statistic.  }
\label{myfit}
  \end{center}
\end{figure*}

The \textit{HST} band is then smoothed using a Gaussian smoothing
kernel normalised to unit area. We then fit the smoothed, scaled
version of the reference {\it HST\/} continuum band to the
delay-corrected longer wavelength continuum band over those epochs
which {\it lie outside of the anomaly\/} identified by Goad
et~al. 2014, and indicated in Figure~\ref{myfit} by the vertical blue
dashed lines (corresponding to epochs JD2456766.1--JD2456829.8). That
is, we implicitly assume that the out of anomaly data indicate what
the normal response of a continuum band to a given amplitude driving
continuum event should look like.  The fit is then optimised using
$\chi^{2}$ minimisation, comparing (as before) only those pairs of
points in both light-curves which lie within $\pm 0.5$~days of one
another, i.e., no where do we interpolate the light-curves, as for
sparsely sampled light-curves, such an approach gives too much weight
to interpolated points in determining the fit.  We optimise the
parameters -- scale factor $\alpha$, offset $K$ and full-width at
half-maximum (FWHM) of the smoothing kernel, by minimising the
residuals between the two light curves in the fitted region. The
results of our fitting procedure are illustrated in
Figure~\ref{myfit}, which shows a fit to the ground-based {\it
  i}$^{\prime}$ light-curve, among the best-sampled ground-based
light-curves (under-sampled light-curves bias the fitting procedure to
find smoother fits). Figure~\ref{myfit} shows the delay-corrected
longer wavelength continuum band (red points) together with the
scaled, and smoothed shortest wavelength \textit{HST} band
($\lambda$1157\AA\/, black points).

Averaging the fluxes of near-contemporaneous (within $\pm$0.5 days)
data pairs taken from the centre of the anomalous period (JD2456777.43
-- JD2456814.97), indicates \textit{that in general} the smoothed and
scaled \textit{HST} light-curve over-predicts the flux in the longer
wavelength continuum bands during this time period.  This effect,
though small, is most apparent in the longer wavelength optical bands
{\it u}$^{\prime}$, {\it g}$^{\prime}$, {\it r}$^{\prime}$, {\it
  i}$^{\prime}$ and {\it z}$^{\prime}$, i.e., the same wavelengths at
which the diffuse continuum contribution to the measured continuum
flux (and so delay signal) is predicted to be strong.  It is also seen
to a lesser extent, in the \textit{Swift} UVW1, UVW2 and U bands,
where the smoothed and scaled \textit{HST} light-curve lies above the
majority of the red points.

In particular, the continuum event centred on day 95 (epoch
JD2456785.7), highlighted in orange in Fig~\ref{fig1}, if present in
the driving continuum light curve, is of sufficient amplitude and
duration to drive an observable response in the more slowly varying
broad emission-lines (Goad et~al. 2016). As we show later, it is also
more than capable of driving a similarly large amplitude response in
the longer wavelength continuum bands (and which vary on even shorter
timescales).  However, this feature is not seen either in the
integrated broad emission-line light-curves reported by Goad
et~al.\ (2016), nor in the longer wavelength continuum light-curves
presented here (e.g., compare the black points and/or blue solid line
with the red points in Figure~\ref{myfit}). Instead, the longer
wavelength continuum bands display reduced amplitude variability {\it
  relative to that predicted\/}, and a reduced flux, during the
anomalous period. While part of this effect may be attributed to
contamination of the \textit{Swift} UV/optical and ground-based
optical broad band filters by broad \textit{emission-lines}, the
longer wavelength {\it i}$^{\prime}$, $I$ and {\it z}$^{\prime}$
continuum bands are free of strong emission lines. \footnote{Though
  thermal emission from dust may contribute to the continuum flux at
  longer wavelengths, the size of the dusty torus in NGC~5548 is large
  ($\approx$ 50 light-days; Suganuma et~al. 2006) even when compared
  to the typical formation radius of major emission-lines. Coupled
  with the relatively small amplitude and uncharacteristically short
  continuum variability timescales present during the AGN~STORM
  campaign, it is unlikely that thermal dust emission contributes
  significantly to the observed continuum variations at longer
  wavelengths.} The degree to which the diffuse continuum from the BLR
affects the continuum band light curves will depend on its
contribution to the measured continuum band flux. We explore this in
future work.

 We repeat this analysis fitting instead to the in-anomaly points
 only. While this reduces the flux deficit between the smoothed
 \textit{HST} light-curve and longer wavelength continuum bands (as
 expected) during the period of the anomaly, the peak centred at day
 95, and apparent in the smoothed \textit{HST} light-curve, is not
 present in the longer wavelength continuum bands. Removal of this
 peak requires a degree of smoothing not warranted by the data, and as
 a consequence results in a poor fit to the out-of anomaly data (due
 to the smaller amplitude variability of the in-anomaly data). By
 expanding the number of fit parameters to include a linear background
 component (solid blue-line), we exclude the possibility that the flux
 deficit in the anomalous region arises due to a tilt in the
 light-curves relative to one another. For completeness, we have also
 performed fits optimised to the pre-anomaly data only and to the
 post-anomaly data only (where possible). In all cases, the excess
 flux in the region of the anomaly predicted by the model remains.

All of these experiments support the notion that there is a flux
deficit in the longer wavelength continuum bands, which generally
should have larger contributions from the BLR diffuse continuum,
relative to that at $\lambda$1157\AA\/, and that there exist
significant differences in the response amplitude of the continuum
light-curves between the start and end of the campaign.

\section{Discussion}

We have used the observed variations of key diagnostic emission lines
to show that the anomalous behaviour observed among the broad emission
lines in NGC~5548 (Goad et~al.\ 2016; Pei et~al.\ 2017) most likely
arises from a change in the strength and shape of the ionising
continuum incident upon the BLR gas. Specifically, the higher energy
continuum band responsible for driving variations in He~{\sc ii}
$\lambda$1640\AA\/ (at energies above 54.4~eV) shows a larger decline
relative to the $\lambda$1157\AA\/ continuum band than the continuum
just shortward of the Lyman limit and responsible for driving
variations in Ly$\alpha$. The ionising SED incident upon BLR clouds is
both \textit{weaker} and \textit{softer} than expected relative to the
$\lambda$1157\AA\/ continuum band during the anomalous period. In the
context of a scenario in which intrinsic changes in the SED took
place, a steepening in the EUV power-law index by $\approx$0.3 would
approximately reproduce the observed unexpected decreases in the
emission-line equivalent widths during the anomalous period (Korista,
Baldwin, \& Ferland 1998; see their Figure~1b).

Supporting evidence for a change in the ionising SED at EUV energies
can be found in simultaneous \textit{Swift}/XRT and \textit{Chandra}
X-ray spectra of NGC~5548 (Mathur et~al.\ 2017). In particular, the
higher S/N \textit{Swift} observations indicate a dramatic
\textit{increase} in the soft X-ray flux (soft excess) at energies of
0.3--2~keV, starting $\approx$10 days prior to the onset of the
anomaly, as determined from the UV broad emission lines (see
Figure~\ref{fig1}). This soft excess reaches a maximum in flux near
the time of the transition into the anomalous period (approximately
near day 74), and then declines thereafter to its pre-anomaly value
some $\approx$40 days later.  While the causal nature between the
observed soft X-ray flare and the anomaly is not yet clear, we believe
a link between the two phenomena to be highly probable. We also note
that the observed soft X-ray emission is along the line of sight, and
our viewpoint may not be representative of that seen by the BLR, i.e.,
there are likely many different sight lines between the continuum
source and BLR clouds which do not intercept our line of sight.
Anomalous behaviour in the intrinsic narrow absorption lines, and in
particular those of high ionization, in NGC~5548 add further
supporting evidence for a change in the SED at soft X-ray energies
along our line of sight (e.g., Kriss et al. 2019, AGN STORM paper~{\sc
  viii} submitted; Dehghanian et al. 2019, AGN STORM, paper~{\sc x},
in press.

An obvious consequence of a weaker and softer EUV continuum is that if
the same gas responsible for producing the broad emission lines also
emits a significant diffuse continuum component, as described in KG01,
then a reduction in the flux of ionising photons incident upon BLR gas
will result in a reduction in the production of both emission lines
and diffuse continuum. Identification of the anomalous behaviour in
the continuum bands therefore provides corroborating evidence for a
significant variable diffuse \textit{continuum} component arising from
BLR gas, and furthermore, affords a means of estimating its overall
contribution. In particular, this represents the strongest
observational evidence yet of the influence of the variable diffuse
continuum at \textit{wavelengths outside of the Balmer jump
  region}. While its influence in the vicinity of the Balmer jump was
suggested by the excess lag in $u^\prime$-band and made clear by the
\textit{HST} lag spectrum of NGC~4593, the presence of the diffuse
continuum from the BLR at longer wavelengths was mainly inferred
(Cackett et~al. 2018). From the point of view of understanding the
central accretion region of AGN, the variable diffuse continuum from
BLR gas represents a nuisance component which must be accounted for in
disk and broad emission-line reverberation mapping experiments, and is
particularly problematic in the longer wavelength optical/near-IR
continuum bands, where the delays and flux contributions are expected
to be generally larger (KG01).

Over the range of measured UV--optical continuum bands, the diffuse
continuum is expected to contribute minimally near the
$\lambda$1157\AA\/ continuum, and generally increases in contribution
toward longer wavelengths, reaching local maxima near the Balmer and
Paschen thresholds. The larger the contribution, the smaller the
inter-band correlation is expected to be.  If this nuisance component
is ignored, the larger measured delays will lead to an overestimate of
the ``size'' of the accretion disk, as has been found from several
recent intensive disk reverberation mapping campaigns of nearby AGN,
including NGC~5548 (Edelson et~al.\ 2015; Fausnaugh et~al.\ 2016;
Edelson et~al.\ 2017; Cackett et~al.\ 2018).  Furthermore, significant
contamination of the optical continuum band by BLR diffuse continuum
emission will lead to an underestimate of the BLR size in optical RM
campaigns, over and above that already imposed by an intrinsically
delayed optical continuum arising from the spatially extended disk.

Models that attempt to explain \textit{all of the inter-band
  UV-optical continuum variations}, without including the expected
contribution from BLR gas, are very likely flawed.  For example,
Gardner \& Done (2017) attribute all of the continuum inter-band
delays to a second reprocessing region, a puffed up Comptonised
region, which they refer to as the EUV torus. This EUV torus
reprocesses hard X-ray emission into lower amplitude more slowly
varying EUV radiation, which in turn, illuminates the outer disk. In
their model the longer wavelength UV--optical continuum inter-band
delays are not dominated by light travel-time effects.  Rather, they
originate within the outer regions of their proposed EUV torus, and
are related to changes in scale height and size as the Comptonised
disk expands and contracts in response to X-ray heating at its inner
edge. While they dismiss contamination of the broad-band UV and
optical continuum by \textit{broad emission lines} (as already
demonstrated by Fausnaugh et~al.\ 2016), missing from their model is
the significant and variable \textit{continuum} emission from the
BLR. Indeed, Fausnaugh et~al. (2016) showed that the enhanced delays
in the longer wavelength $R$ and $r^{\prime}$ bands could reasonably
be explained in terms of contaminating broad H$\alpha$ ($\approx$20\%
and $\approx$15\% respectively), while those in the vicinity of the
little blue bump ($U$ and $u^{\prime}$) could be attributed to a
significant ($\approx$19\%) diffuse continuum component from BLR gas
with similar delays as the Balmer lines.

The substantial drops in delay longward of the Balmer and Paschen
jumps, a strong prediction of photoionisation model calculations and a
key diagnostic of a significant diffuse continuum component (KG01;
Lawther et~al.\ 2018), is clearly visible in their data (Edelson
et~al.\ 2015; Fausnaugh et~al.\ 2016). The best evidence to date for
significant Balmer and Paschen jumps are seen in the intensive
multi-wavelength spectro-photometric monitoring data for NGC~4593
(Cackett et~al.\ 2018). Indeed, McHardy et~al.\ (2018) identify the
extended tails in the response functions for NGC~4593 with reprocessed
continuum emission from the surrounding gas (i.e., the BLR).

Starkey et~al. (2017) simultaneously fit a linearised $\alpha$-disk
model to the 19 \textit{HST/Swift} and ground-based continuum data
taken as part of the AGN STORM monitoring campaign on NGC~5548.  They
find that the standard $\alpha$-disk model $T(r)\propto r^{-\alpha}$,
with $\alpha$ fixed at 3/4, is a reasonably good fit to the
UV--optical continuum light-curves, except for $u^{\prime}$, where the
model tends to {\em lead the data and is more highly variable\/}. To
fit the model, they require consistently larger rescaling of the error
bars at longer wavelengths, which may indicate that the longer
wavelength variations cannot adequately be described by a simple
linearised echo model. These findings we argue are further evidence
for contamination of the delay signal by diffuse continuum originating
in the BLR.

In summary, the detection of the anomaly in the UV and optical
continuum bands, coupled with the increased delays over and above that
expected from a simple disk reprocessing scenario, and the presence of
significant delay changes either side of the Balmer and Paschen jumps,
we suggest provides the first clear evidence of a significant
contribution to the UV-optical continuum from broad emission line
region gas.

\section{Conclusions}

A standard assumption in RM is that the observed continuum, and in
particular the far--UV continuum, is a suitable proxy for the driving
continuum at ionising energies. However, during the 2014 AGN~STORM
monitoring campaign of NGC~5548, this assumption was violated, as
evidenced by the decorrelation between the strong UV broad emission
lines and the shortest accessible UV continuum band approximately
mid-way through the campaign, here referred to as the anomalous
period.  However, we have shown that by using key diagnostic emission
lines (Ly$\alpha$ and He~{\sc ii}) we can infer a proxy for the
variable driving ionising continuum incident upon BLR clouds.  For
NGC~5548, these key diagnostic emission lines indicate a weakening and
softening of the EUV continuum ($\gtrsim$30\% in the Lyman continuum
and $\gtrsim$40\% in the 4 Rydberg continuum) relative to the shortest
available \textit{HST}/COS continuum band ($\lambda$1157\AA\/) during
the period of anomalous behaviour exhibited by the strong broad UV and
optical emission lines. In addition, most of the ionising continuum
was apparently much less variable than was the UV continuum during the
anomalous period. The derivation of an appropriate proxy for the
behaviour of the time-variable driving ionising continuum is crucial
first step towards the recovery of the broad emission-line response
functions in this source.

We next compared the level of correlation between the $\lambda$1157
continuum fluxes and the fluxes within all other UV-optical continuum
bands, \textit{both outside and during }the anomalous period.  We
found a general trend that the Spearman rank correlation coefficient
diminished with increasing wavelength of the continuum band, but also
that the correlation during the anomalous period was significantly
depressed relative to that outside this time period,
\textit{particularly within wavelength bands where we expect the
  contribution from the continuum emanating from the BLR is greater}.
This is what we would expect if the BLR is contributing to the
continuum flux measurements \textit{and} if the BLR saw a weaker and
less variable ionising EUV continuum during the anomalous period.  We
also compared scaled, smoothed, and time-delayed versions of the
$\lambda$1157 continuum light curve to each of the other UV-optical
continuum bins as another means of exploring the likely decorrelation
of the EUV continuum from the observed UV-optical continuum, using the
potential presence of the BLR diffuse continuum within the continuum
bands as a marker of this decorrelation in flux and variability.

Importantly, we find corroborative evidence for anomalous behaviour in
{\it virtually\/} all of the longer wavelength (relative to
$\lambda$1157\AA\/) UV and optical continuum bands. This we suggest
indicates a significant contribution to the variable UV--optical
continuum emission from the diffuse continuum emitted by BLR
clouds. This has a direct impact on the physical interpretation of the
measured inter-band continuum delays, and by inference, the size and
temperature dependence of the accretion disk.







\appendix

\bsp
\label{lastpage}

\end{document}